\documentclass[conference]{IEEEtran}
\IEEEoverridecommandlockouts

\usepackage{cite}
\usepackage{amsmath,amssymb,amsfonts}
\usepackage{algorithmic}
\usepackage{graphicx}
\usepackage{textcomp}
\usepackage{xcolor}
\usepackage{subcaption}
\usepackage{balance}

\def\BibTeX{{\rm B\kern-.05em{\sc i\kern-.025em b}\kern-.08em
    T\kern-.1667em\lower.7ex\hbox{E}\kern-.125emX}}
\begin{document}

\title{Automatic Platform Configuration and Software Integration for Software-Defined Vehicles
}

\author{Fengjunjie Pan$^1$, Jianjie Lin$^1$, and Markus Rickert$^2$
	\thanks{$^1$ F. Pan, J. Lin are with Robotics, Artificial Intelligence and Real-Time Systems, School of Computation, Information and Technology, Technical University of Munich, Munich, Germany. \{panf, jianjie.lin\}@in.tum.de}%
	
	\thanks{$^2$ M. Rickert is with Multimodal Intelligent Interaction, Faculty of Information Systems and Applied Computer Sciences, University of Bamberg, Bamberg, Germany. markus.rickert@uni-bamberg.de}%
}

\maketitle

\begin{abstract}
In the automotive industry, platform configuration and software integration are mostly manual tasks performed during the development phase, requiring consideration of various safety and non-safety requirements. This manual process often leads to prolonged development cycles and provides limited flexibility.
This paper introduces a novel approach to automate platform configuration and software integration for software-defined vehicles (SDVs), shifting these activities from the development phase to runtime.
Our approach features an integration manager that combines model-based methods and virtualization technologies to generate and execute deployment plans. 
By leveraging model-based systems engineering (MBSE), our method automatically generates platform configuration and software integration plans, which are then converted into deployment-ready formats using code generation techniques.
Utilizing virtualization and container orchestration technologies, the proposed system enables dynamic and flexible resource allocation while ensuring compliance with safety requirements. 
Communication between the development and runtime platforms is facilitated via a REST API.
A proof of concept was implemented on a simulated SDV platform with the Intel Whiskey Lake Board. This demonstration showcases the integration manager on an SDV with a central computer, highlighting the potential to shorten development cycles and adapt to diverse vehicle configurations.

\end{abstract}

\section{Introduction}
The development of modern vehicles is a complex process that often takes several years to complete. 
Traditionally, these development cycles are managed by human engineers. 
For vehicular software integration, system integrators must ensure that all safety and non-safety requirements are met by configuring the hardware platform and generating software integration plans.
Software-defined vehicles (SDVs) are characterized by features primarily enabled through software, with each vehicle's software configuration potentially differing due to owner customization.  
It is nearly impossible to pre-determine a universally correct configuration. 
Therefore, the current integration procedure requires a significant revolution.

This paper proposes shifting integration activities from development time to the runtime, allowing vehicles to automatically configure their execution environments and install or update software components based on user preferences. 
Similar features of application and service management are found in cloud systems~\cite{Tosatto2015} and mobile phones~\cite{Beimborn2013}. 
While these platforms offer great flexibility during runtime, the software integration on them does not typically concern about various requirements, such as isolation among applications with different safety levels. 
This results in fewer constraints during integration activities.  
Furthermore, these platforms often provide sufficient computational resources for their software, leading to limited discussion on their planning and configuration in the context of resource allocation.
In the automotive industry, Over-The-Air (OTA) \cite{Chowdhury2018} technology can manage applications on the customer side. 
However, decisions regarding the integration of safety-critical software are still conducted during the design phase primarily, with OTA handling only installation and updates. 
No current approach allows a flexible software integration of both safety-critical and non-safety software on SDVs.

\begin{figure}[t]
	\centering
	\vspace{0.1cm}
	\includegraphics[width=0.47\textwidth]{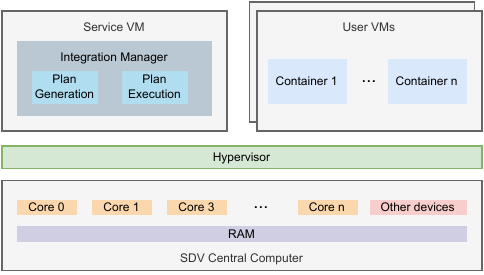}
	\caption{Architecture of central computer on SDVs}
	\label{fig:sdv}
\end{figure}

In this work, we focus on automated platform configuration and software integration for SDVs with a centralized architecture.
In our proposed approach, we design an integration manager that combines model-based system design with virtualization-based deployment methods on the vehicle's central computing platform (CCP) (Fig.~\ref{fig:sdv}).
An optimization engine is embedded in the integration manager to enable online decision-making and optimization of platform configuration and software integration. 
This process of finding integration plans is based on the model-based design space exploration (DSE) method presented in our previous work \cite{Pan2023}. 
In addition, code generation is employed to create executable files based on the integration plan.
We further utilize virtualization technologies for the deployment to guarantee freedom of interference among software, ensuring that the failure of a single application does not affect the execution of others.
Hardware virtualization (hypervisors) creates isolated partitions on the same hardware to host applications with varying requirements.  
Within each partition, OS-level virtualization (containers) separates individual applications. 
Additionally, we introduce the infrastructure as code concept, using tools like Terraform \cite{Brikman2022} with Kubernetes extensions \cite{Burns2018} to manage, execute, and monitor deployment tasks. 
The REST API \cite{Masse2011} is used to transmit and trigger integration requests on the SDV platform.

The feasibility of this approach is demonstrated through a simulated SDV platform using the Intel Whiskey Lake board \cite{WL10}, the ACRN hypervisor \cite{ACRN}, the lightweight Kubernetes K3S \cite{K3S}, and Terraform. 
A Java runnable program, based on our previous work \cite{Pan2023}, was created for automated design space exploration (DSE). 
We further utilize Eclipse Acceleo~\cite{acceleo} for generating deployment files.
Eclipse Jersey has been used to send integration requests from the development PC to the SDV platform. 
The results indicate the potential of automated platform configuration and software integration to shorten vehicle development cycles. 
At the end of this paper, we further discuss the potential extension of the current integration manager concept.

\section{Related Work}
The automatic platform configuration and software integration in SDVs includes two primary aspects: system planning and deployment execution.

Automated system planning has been discussed in various researches incorporating model-based and formal methods. MechatronicUML~\cite{Dziwok2014} supports creating system modeling and solving automotive resource allocation based on predefined constraints. This framework also facilitates C code generation and allows for simulation of designed models using tools such as MATLAB Simulink.
Eclipse APP4MC~\cite{Hottger2017} focuses on scheduling analysis and performance simulation in multi-core software systems through a model-based development approach.
AutoFocus3~\cite{Aravantinos2015} presents s system modeling method with predefined meta information and constraint template. It supports verification, design space exploration for the development of embedded systems and is capable of generating hardware-specific code.

While these works primarily focus on automated system planning for a flexible SDV platform, the concept of deployment execution must also be addressed. Many approaches for software deployment have been proposed for personal computers.
Package managers, e.g., Ubuntu’s Advanced Packaging Tool (APT), automate the processes of installing, upgrading, configuring, and removing software, managing dependencies to prevent dependency issues.
Flatpak~\cite{flatpak} allows users to install applications in separate environments on Linux systems. It provides developers control over dependencies and updates, using sandboxing technologies for enhanced security and cross-distribution software distribution.
Containers, e.g., Docker~\cite{docker}, package code and dependencies, ensuring consistent execution across environments. Container images are lightweight, standalone packages containing all necessary information to run an application, including code, runtime, system tools, libraries, and settings.
In the automotive domain, the Scalable Open Architecture for Embedded Edge (SOAFEE) project~\cite{Matt2021} is a collaborative initiative to utilize cloud-native virtualization methods for automotive applications. 

Our previous work has discussed the SDV's system planning and deployment execution separately. In \cite{Pan2023}, we proposed a flexible and system-independent approach for model-based design space exploration, transforming user-defined model information and constraints into an optimization problem.
In \cite{Wen2023}, we explored automotive software deployment using hypervisors and containers, discussing the performance impact of such technologies for running automated applications.
This work aims to bridge the gap between model-based system planning and virtualization-enhanced deployment execution, facilitating an automated approach for the platform configuration and software deployment of SDVs.
\section{Preliminary}
\begin{figure*}[t]
	\centering
	\vspace{0.1cm}
	\begin{subfigure}{\textwidth}
		\includegraphics[width=1\textwidth]{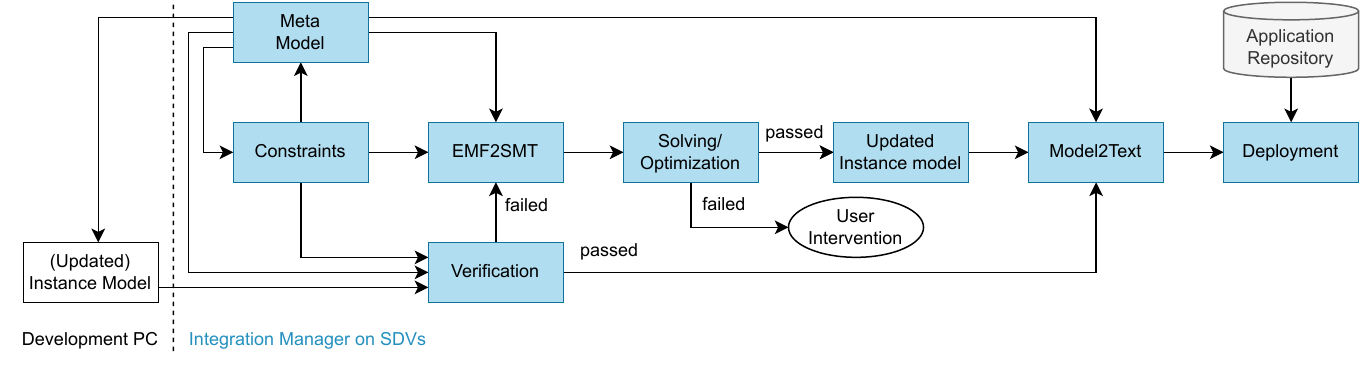}
		\caption{An online approach performed on SDVs}\label{online}
	\end{subfigure}
	\begin{subfigure}{\textwidth}
		\includegraphics[width=\textwidth]{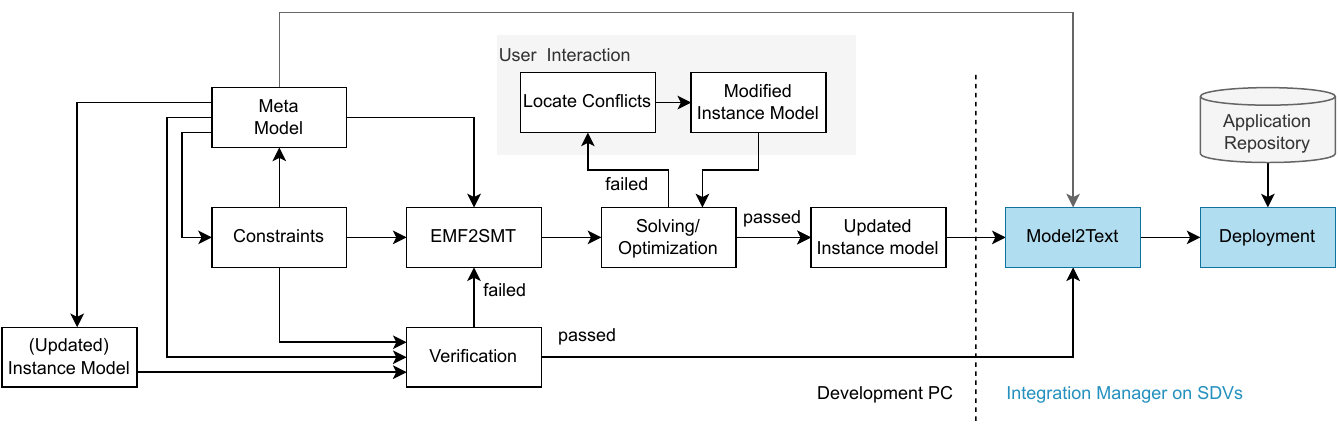}
		\caption{A design time approach}\label{offline}
	\end{subfigure}
	\caption{Automated platform configuration and software integration workflow}
	\label{fig:workflow}
\end{figure*}

\subsection{Model-based design space exploration}
Model-based engineering is widely used for system design in the automotive and aviation industries. It leverages formal models and constraints to describe systems and the requirements that need to be fulfilled during the design process. Meta-modeling is an essential concept in model-based approaches. A meta-model provides a high-level abstraction of systems, including system component types, attribute types, and relationship types. Based on the meta-model, a concrete system can be instantiated in an instance model, which contains a detailed description of the system. Additionally, formal constraints can be defined based on types in the meta-model to constrain the concrete system objects in instance models.

In \cite{Pan2023}, a model-based design space exploration approach was discussed to solve the automotive resource allocation problem, where all model information (meta-model, instance model and constraints) is transformed into optimization problems and solved accordingly. This procedure has been integrated into our workflow (Fig.~\ref{fig:workflow}) to identify the desired platform configuration and software integration plan.

\subsection{Hardware Virtualization}
Hardware virtualization allows a single machine to create multiple simulated environments, known as virtual machines (VMs). This process involves abstracting physical resources, isolating them, and distributing them to different VMs~\cite{Desai2013}, typically through a hypervisor. 

There are two main types of hypervisors. Type 1 hypervisors run directly on the host's hardware to manage user operating systems, with VM resources directly mapped to the hardware to ensure strong isolation among VMs. Type 2 hypervisors run as a software layer on a host operating system (OS), abstracting the guest OS from the host OS. In this case, VM resources are scheduled based on the host OS and then executed on the hardware. This approach makes VM configuration can be easily managed via the host OS. The distinction between these hypervisor types is becoming less clear, as many modern hypervisors support direct hardware access while also offering host system capabilities for VM configuration. In this work, we utilize the ACRN hypervisor~\cite{Li2019} as a proof of concept. ACRN is a flexible and lightweight hypervisor with a focus on real-time performance and safety. It provides a host VM to manage user VMs.

\subsection{OS-level Virtualization}

OS-level virtualization, commonly known as containerization, allows applications running in isolated environments on the same operating system~\cite{Turnbull2014}. Unlike VMs, which require a full operating system for each instance, containers share the host machine's OS kernel. This approach leads to more efficient resource utilization and reduced overhead.

A container image is an executable package of software that includes everything needed to run an application, such as code, runtime, system tools, system libraries, and settings. These images become containers when run on a container engine, such as Docker. The isolation provided by containers ensures that software will always run the same way, regardless of the underlying infrastructure. It brings flexibility and reliability in software deployments.

A container orchestrator, such as Kubernetes, is a tool for managing the lifecycle of containers~\cite{Burns2022}. It can manage containerized applications across multiple hosts, providing basic mechanisms for deployment, maintenance, and scaling of applications. 
In this project, we utilize a lightweight distribution of Kubernetes, K3S~\cite{K3S}, as part of the proof of concept. 

\section{Approach}

This study aims to develop an automatic platform configuration and software integration process for SDVs. Traditionally, car manufacturers have had to recall sold vehicles to update their software, requiring expert system engineers to manage the integration. The proposed workflow (Fig.~\ref{fig:workflow}) works on vehicle platforms utilizing visualizations (Fig.~\ref{fig:sdv}). It introduces automated steps for both online and design time approaches, allowing platform configuration and software integration to be analyzed on both development environment and the vehicle runtime, and performed on the vehicle. Both approaches share similar procedures.

The online workflow (Fig.~\ref{online}) begins with sending an instance model as an integration request from the development PC to the SDV. 
A instance model typically represents a  concrete vehicle system containing hardware resources and software. 
Ideally, a complete instance model should specify system configurations including resource allocation decisions so that the SDV can be directly deployed.
However, in our workflow, the instance model does not necessarily need to be complete and may contain only partial or no resource allocation decisions, as the automated approach for integration manager will verify the system configuration and generate the deployment plan if necessary.
For instance, when adding, deleting, or updating applications, the instance model can be modified accordingly by the system integrator or through automated scripts and then sent via a REST API to the integration manager, which resides in the vehicle's CCP, for automated processing.

After the integration manager receives the instance model, it verifies it against predefined constraints using a verification engine. These constraints and the meta-model are established during the development phase and cannot be modified at runtime. 
If the system configuration is complete, deployment files will be generated via model-to-text code generation techniques for further processing. If the system configuration is incomplete, a dedicated algorithm EMF2SMT converts the constraints to SMT format for the optimization engine in the integration manager to solve, following the approach discussed in previous research \cite{Pan2023}.
The resulting system configuration is produced as a deployment-ready complete instance model, which is then transformed into deployment files. For unsolvable issues, manual intervention by a platform integrator may be required. 

In the SDV platform, the CCP is divided into isolated partitions (VMs) using hypervisors (Fig. 1). The integration manager is located within a specific partition called the Service VM, while applications in the vehicle are located in partitions called User VMs. OS-level virtualization (containers) is used within each User VM to execute and isolate applications. These applications are packaged into container images and stored in a container registry.
Once the integration plan has been generated, the integration manager triggers the hypervisor module to properly configure the User VMs and uses a container orchestrator to download and deploy applications on them. 

A similar integration approach can be applied as a design-time approach (Fig.~\ref{offline}), where optimization is performed on the development PC. In this offline approach, human engineers have the possibility to inspect conflicts if the expected integration cannot be planned, and modify the instance model recursively to meet the integration target system. After the design-time analysis, a complete instance model is sent to the integration manager on SDVs to trigger the integration activity.

The online integration approach is performed on the SDV, which typically has limited computational resources and might therefore be restricted to relatively small-scale integration tasks. The design-time approach, where the analysis of integration is mostly done on the development PC, is suitable for handling large-scale problems. The system architecture of the integration manager is introduced in Section~\ref{sec:manager}.

\section{Integration Manager}\label{sec:manager}
\begin{figure}[t]
	\centering
	\vspace{0.1cm}
	\includegraphics[width=0.47\textwidth]{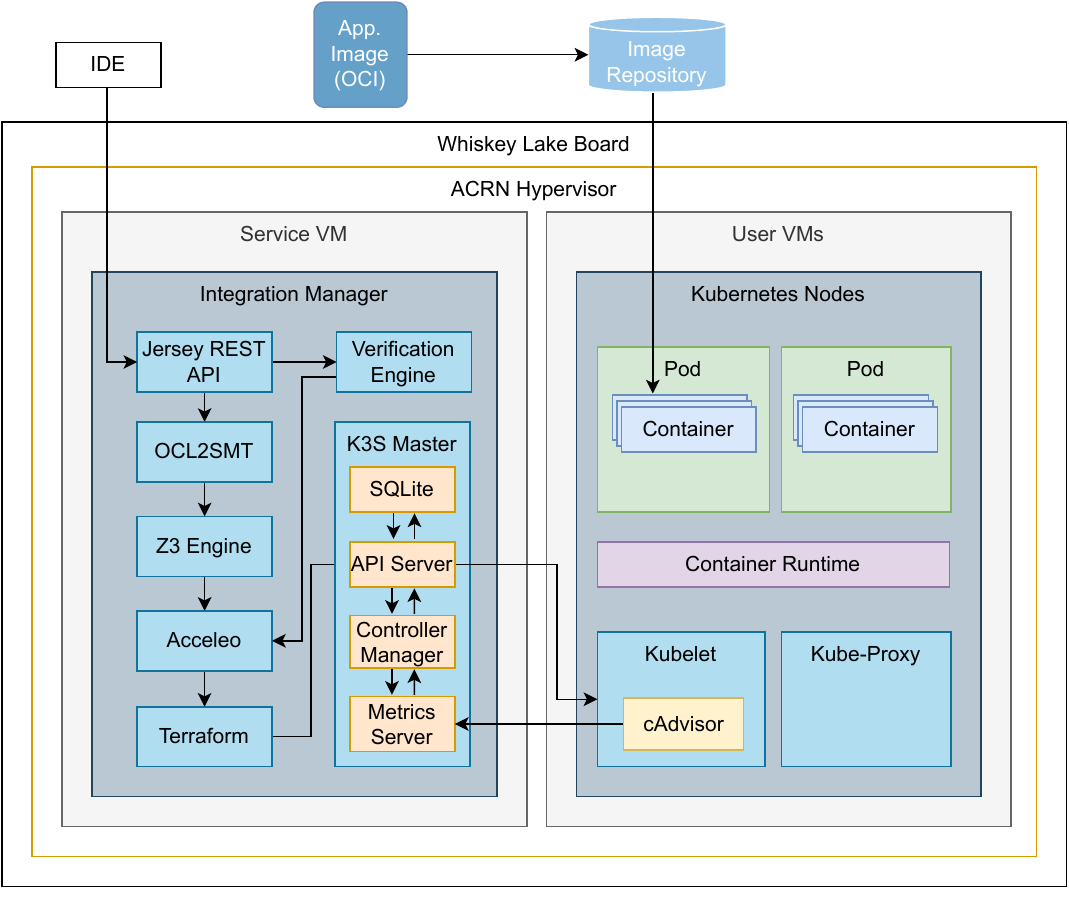}
	\caption{PoC-based integration manager architecture}
	\label{fig:imanager}
\end{figure}

The integration manager focuses on automatic platform configuration and software integration for software-defined vehicles. Fig.~\ref{fig:imanager} presents a reference architecture of the integration manager based on our proof of concept (PoC) setup (Section~\ref{sec:poc}).

\begin{figure}[t]
	\centering
	\vspace{0.1cm}
	\includegraphics[width=0.47\textwidth]{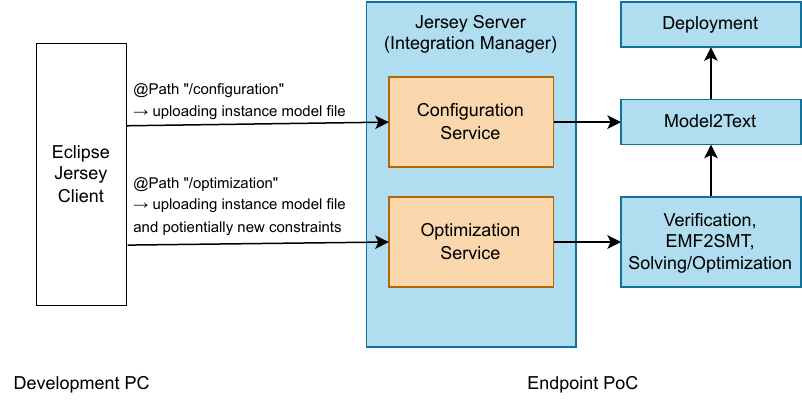}
	\caption{Communication between IDE and integration manager}
	\label{fig:restapi}
\end{figure}

The integration process begins within an integrated development environment~(IDE), such as Eclipse, functioning as the REST API client. 
Through the IDE, system integrator can create and send model files as requests to the server for software update or installation.
The REST API server, implemented via the Eclipse Jersey framework \cite{Jersey}, resides on the CCP to receive requests from the client. 
Fig.~\ref{fig:restapi} illustrates the communication between the IDE on the development PC and the integration manager on the PoC CCP.
We implemented two endpoints. The configuration service has been designed for the design time integration approach (Fig.~\ref{offline}). It receives the instance model with a complete configuration, which then triggers the generation of executable files for deployment.
The optimization service receives incomplete instance model and potentially updated constraints. It triggers the online integration approach (Fig.~\ref{online}), which includes the automated solution generation process and subsequently generates deployment files.

The vehicle's CCP is virtualized using a hypervisor. In this work, we employed the ACRN hypervisor~\cite{ACRN}, which includes a Service VM and multiple isolated virtual machines. The Service VM manages and launches user VMs for deploying applications using container orchestration tools.
Upon the REST API sever in Service VM receiving the files, the verification engine initiates the validation process for the instance model based on a given meta-model and constraints.  
Incomplete configurations are further processed by the optimization engine. However, the optimization engine cannot directly process the configuration file. Therefore, the EMF2SMT module converts models and constraints into optimization models in SMT format, for which the Z3 engine can be used for solution finding and optimization. 
The solved configuration is then translated the Terraform deployment file via Acceleo.

Terraform works as orchastrator for the integration tasks. It coordinates with K3S~\cite{K3S}, a lightweight Kubernetes, to execute the deployment. Terraform compares the desired system configuration against the current state and move the system to desired state. K3S contains multiple services for container orchestration~\cite{K3SArch}. In Kubernetes, SQLite is used as a database engine to store data accessible by a cluster of machines. The Kubernetes API server enables querying and manipulating the state of API objects within Kubernetes. The Kubernetes controller manager, a daemon embedding control loops, monitors the shared state of the cluster and makes necessary adjustments to achieve the desired state. The Kubernetes metrics server collects resource metrics from cAdvisor within kubelets, which analyze and expose resource usage and performance data from running containers. Kubelet, called by the API server, manages the container runtime to create different pods on the Kubernetes node. The Kubernetes node uses Kube-proxy, a network proxy that maintains network rules, facilitating communication to pods from both internal and external network sessions.

This system architecture supports a flexible and automated approach to platform configuration and software integration in software-defined vehicles, enhancing efficiency and reducing the need for manual intervention.
\section{Proof of Concept}\label{sec:poc}
For the demonstration of the proposed approach, we utilized the Whiskey Lake board~\cite{WL10} as the CCP of an SDV. The board is equipped with heterogeneous resources, including 8 CPU cores, an integrated GPU (iGPU), and 8 GB of RAM.
We selected eight applications from OpenVINO \cite{Openvino} (security barrier camera, object detection), ROS \cite{Ros} (roscore, rosbag, SLAM node, and Rviz), and well-known benchmark tools (glxgears and stress tool) as the example software set. 
The applications are categorized as either safety or non-safety, and either CPU or GPU-related applications.  Each application is built as an OCI container image~\cite{OCI} and updated to a local image repository.

To provide separate execution environments for applications with different requirements, VMs with the Ubuntu operating system are created by the ACRN hypervisor.  
Due to ACRN's limited flexibility for runtime configuration, we predefined VMs for deployment.  
We created one Service VM and two User VMs. The Service VM is equipped with 2 cores and 4 GB of RAM. It hosts the integration manager and is responsible for the configuration of user VMs. User VM 1 is equipped with 4 cores and 8 GB of RAM and is designed as a safety VM to host safety-critical applications. 
User VM 2 is allocated 2 cores and 4 GB of RAM and is used to execute non-safety-critical applications. 
All VMs share the same iGPU and Ethernet port.

\begin{figure}[t]
	\centering
	\vspace{0.1cm}
	\includegraphics[width=0.49\textwidth]{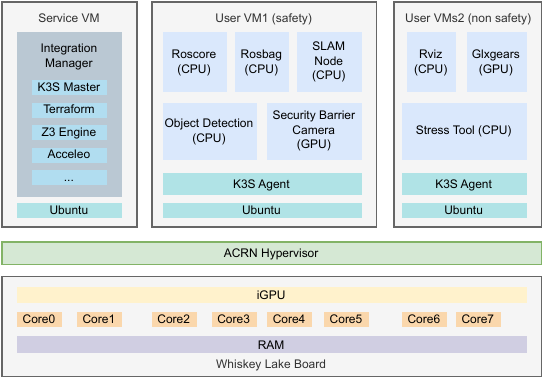}
	\caption{Integration of applications on the whiskey lake board via integration manager}
	\label{fig:SDVintegration}
\end{figure}

The automated integration follows the constraints defined in \cite{Pan2023}.
As both online and offline approaches follow similar principle,
we demonstrated a two-step integration scenario with pure online approach. In the first step, OpenVINO and ROS applications are intended to be installed. In the second step, glxgears and the stress tool should be integrated into the running system.
Each step triggered the complete process of the integration manager. In the first step, an instance model containing the target applications is sent to the integration manager without specifying which VMs should host them. The integration manager calculates the integration decision and then executes it. In the second step, we added new applications to the instance model containing the integration solution from the first step. The integration manager resolves the resource allocation problem again without altering the existing deployment and deploys the new applications.
The resulting SDV system in the end is presented in Fig.~\ref{fig:SDVintegration}

\begin{figure*}[t]
	\centering
	\vspace{0.1cm}
	\begin{subfigure}{0.47\textwidth}
		\includegraphics[width=\textwidth]{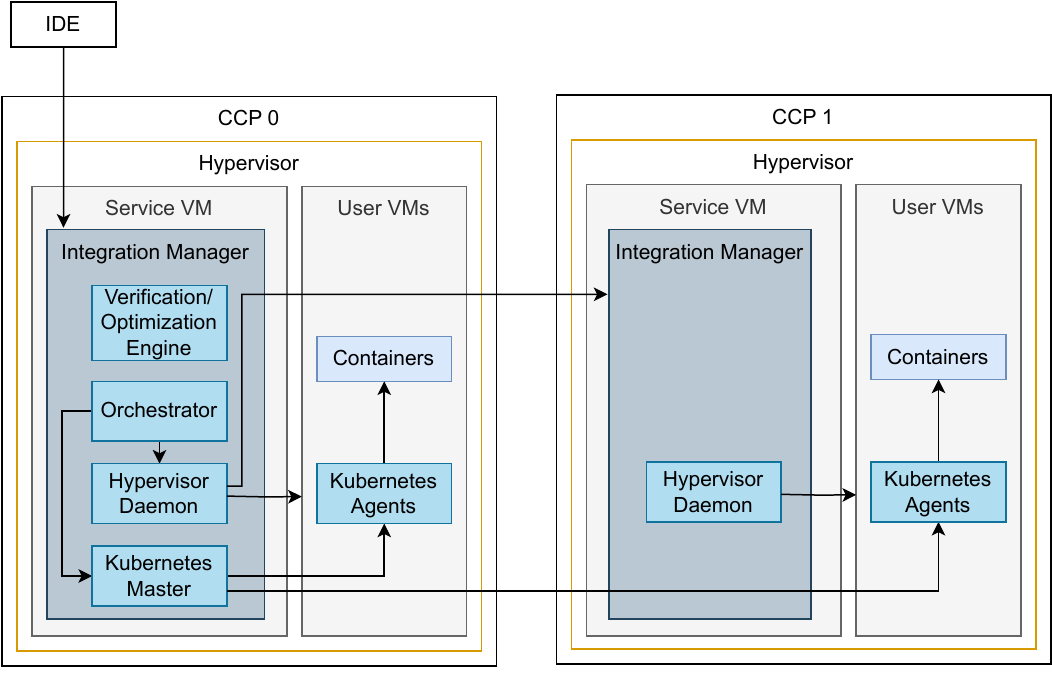}
		\caption{Integration manager for multiple CCPs}\label{fig:multiCCP}
	\end{subfigure}
	\hspace{0.6cm}
	\begin{subfigure}{0.47\textwidth}
		\includegraphics[width=\textwidth]{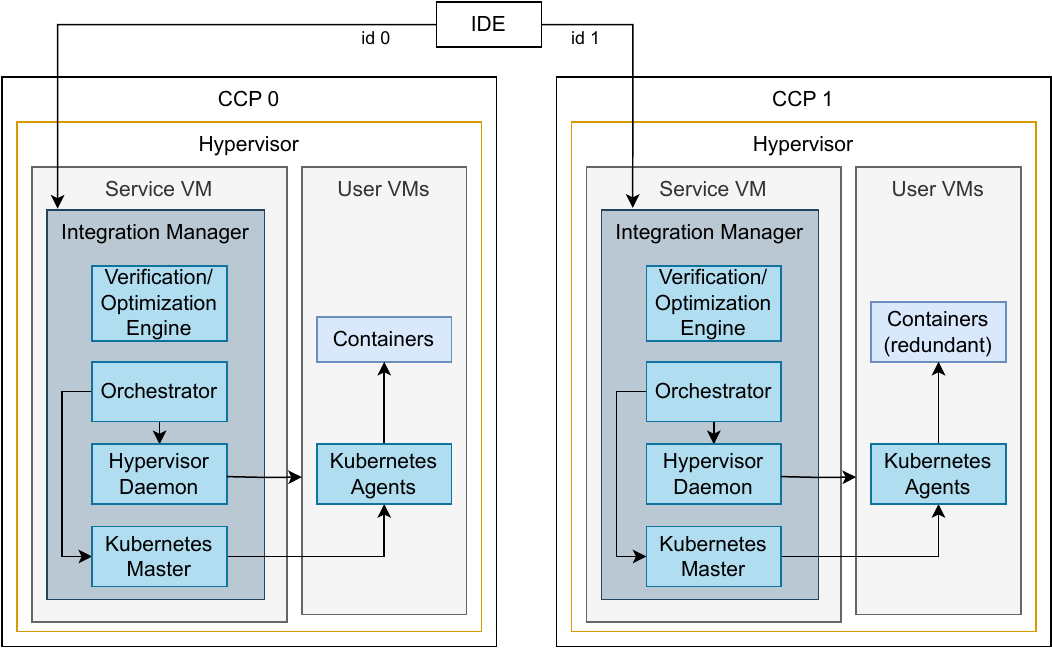}
		\caption{Integration manager for redundant CCPs}\label{fig:redundantCCP}
	\end{subfigure}
	\caption{Extended concept of integration manager}
\end{figure*}

Compared to the traditional manual integration approach, the integration manager simplifies the process with a one-click operation, significantly easing integration and providing flexibility to the SDV system.

\section{Extended Concept of Integration Manager}
In previous sections, the concept of an integration manager was mainly discussed for a single CCP of SDVs.
However, in reality, an SDV can be equipped with multiple CCPs for consideration of computational power or redundancy requirements.
This section discusses the potential extension of the current integration manager.

One possible extension involves introducing a hierarchy of integration managers with a master manager receiving the integration requests and interacting with multiple integration manager instances across all available CCPs (Fig.~\ref{fig:multiCCP}). Each CCP maintains a similar base architecture, where containers run in different VMs created by the hypervisor. In the proposed concept, one CCP serves as the master, equipped with a single verification and optimization engine to verify and optimize the configuration. This master CCP employs a single orchestrator to manage orchestration (VMs and containers) among all CCPs. Each user VM is equipped with a Kubernetes agent to execute containers. For VM configurations, each CCP is equipped with a separate hypervisor daemon.

In automotive systems, redundant components are crucial for increasing reliability. Fig.~\ref{fig:redundantCCP} illustrates the concept of extending the integration manager for redundant computation channels. This concept can be applied to both homogeneous and heterogeneous hardware/software setups. In this configuration, each CCP operates standalone with an identical and complete architecture stack. Each CCP has its own instance of a REST server, verification and optimization engine, integration orchastrator, Kubernetes master, and hypervisor daemon.  In case of failure in one CCP, a redundantly deployed application in other CCPs can take over. Different configuration schemes for each CCP can be defined manually or by setting different optimization goals. For applications requiring redundant deployment, it should be determined which application instance is active.
\section{Conclusion}
This work investigates automatic platform configuration and software integration for SDVs. In SDVs, we propose to use both hardware and OS-level virtualization to build up the architecture of SDV’s CCP. 
VMs established by the hypervisor provide separate execution environments for individual applications with varying requirements.  Applications run in containers, allowing for more flexible deployment compared to standard strategies.

We introduce an automated in-vehicle workflow implemented as an integration manager for platform configuration and software integration. This integration manager enables systems to analyze, generate, and execute individual integration plans. Within the integration manager, we employ an automated DSE approach utilizing optimization solvers to solve resource allocation problems during software integration. Container orchestration technologies manage the integration execution, facilitating the flexible redefinition of system configurations and application deployments, thereby enhancing the overall efficiency and adaptability of SDVs.

We provided a proof of concept demonstrating the software integration capabilities of integration managers in a simulated SDV environment. Future work will focus on developing a more comprehensive demonstration using realistic automotive hardware and software to further showcase and refine the proposed approach.

\bibliographystyle{IEEEtran}
\bibliography{ref}
\balance
\end{document}